\documentclass[journal=jacsat,manuscript=article]{achemso}
\usepackage{graphicx}% Include figure files
\usepackage{dcolumn}% Align table columns on decimal point
\usepackage{bm}% bold math
\usepackage[section]{placeins}
\usepackage[utf8]{inputenc}
\usepackage{mathptmx}
\usepackage{multirow}
\usepackage{braket}
\usepackage{amssymb}

\usepackage[english]{babel}
\usepackage{xcolor}% Include colors for document elements
\colorlet{RED}{red}
\colorlet{BLUE}{blue}
\usepackage{dcolumn}% Align table columns on decimal pointhttps://v1.overleaf.com/17481932tmwjwzytcbcq#
\usepackage{bm}% bold math
\usepackage[version=4]{mhchem} % HvD: for general chemical formula like \ce{(TiO2)n}
\usepackage{acronym}
\usepackage{adjustbox}
\usepackage{float}
\usepackage{tikz}
\usepackage{microtype} % fixes many of the overflow hbox errors
\usepackage{algorithm}
\usepackage{xcolor}
\usepackage{algpseudocode}
\usepackage{comment}

\newcommand{\onlinecite}[1]{\hspace{-1 ex} \nocite{#1}\citenum{#1}}

\usetikzlibrary{calc,shapes.geometric,decorations.pathmorphing,patterns}

\definecolor{background-color}{gray}{0.98}
\title{
Integrated Software/Hardware Execution Models for High-Accuracy Methods in Chemistry
}

\author{Nicholas Bauman}
\email{nicholas.bauman@pnnl.gov}
\affiliation{
  Physical Sciences Division, 
  Pacific Northwest National Laboratory, Richland, Washington, 99354, USA
}

\author{Ajay Panyala}
\email{ajay.panyala@pnnl.gov}
\affiliation{Advanced Computing, Mathematics, and Data Division, Pacific Northwest National Laboratory, Richland, Washington 99354, USA}

\author{Libor Veis}
\email{libor.veis@jh-inst.cas.cz}
\affiliation{
J. Heyrovsky Institute of Physical Chemistry, v.v.i., Czech Academy of Sciences, Prague, Czechia
}

\author{Jiri Brabec}
\email{jik.jiri@gmail.com}
\affiliation{
J. Heyrovsky Institute of Physical Chemistry, v.v.i., Czech Academy of Sciences, Prague, Czechia
}

\author{Paul Rigor}
\email{paul.rigor@pnnl.gov}
\affiliation{
Center for Continuum Computing, Pacific Northwest National Laboratory, Richland, Washington 99354, USA
}

\author{Randy Meyer}
\email{randymeyer@micron.com}
\affiliation{
Micron Technology Inc., Richardson, Texas 75081, USA
}

\author{Skyler Windh}
\email{swindh@micron.com}
\affiliation{
Micron Technology Inc., Richardson, Texas 75081, USA
}

\author{Craig Warner}
\email{craigwarner@micron.com}
\affiliation{
Micron Technology Inc., Richardson, Texas 75081, USA
}

\author{Tony Brewer}
\email{tbrewer@micron.com}
\affiliation{
Micron Technology Inc., Richardson, Texas 75081, USA
}

\author{Karol Kowalski}
\email{karol.kowalski@pnnl.gov}
\affiliation{
  Physical Sciences Division, 
  Pacific Northwest National Laboratory, Richland, Washington, 99354, USA
}
\affiliation{Department of Physics, University of Washington, Seattle, Washington 98195, USA}

\begin{document}

\begin{abstract}
The effective deployment and application of advanced methodologies for quantum chemistry is inherently linked to the optimal usage of emerging and highly diversified computational resources. This paper examines the synergistic utilization of Micron memory technologies and Azure Quantum Element cloud computing in Density Matrix Renormalization Group (DMRG) simulations leveraging coupled-cluster (CC) downfolded/effective Hamiltonians based on the double unitary coupled cluster (DUCC) Ansatz. We analyze the performance of the DMRG-DUCC workflow, emphasizing the proper choice of hardware that reflects the numerical overheads associated with specific components of the workflow. We report a hybrid approach that takes advantage of Micron CXL hardware for the memory capacity intensive CC downfolding phase while employing AQE cloud computing for the less resource-intensive DMRG simulations. Furthermore, we analyze the performance of the scalable ExaChem suite of electronic simulations conducted on Micron prototype systems. 
\end{abstract}

\maketitle

\section{Introduction}
The development of electronic structure methods to describe complicated processes in chemistry, typically encountered in areas such as catalysis, rational materials design, separation science, biological chemistry, and heavy element chemistry (to name only a few areas of applications), is contingent upon our ability to describe and capture various types of correlation effects. A significant methodological inroads have been achieved in several areas associated with Density Functional Theory (DFT)\cite{teale2022dft}, Density Matrix Renormalization Group (DMRG)\cite{White1992,verstraete2023density}, Configuration Interaction (CI) methods \cite{casanova2009double,levine2020casscf,jimenez2015cluster,schriber2016communication,vogiatzis2017pushing,schriber2018combined,tubman2020modern,mahajan2022selected,abraham2020selected,braunscheidel2024accurate},  various formulations of Coupled Cluster (CC) theory \cite{evangelista2018perspective,Brabec2012,ye2019bootstrap,lee2019excited,vidal2019new,calvin2020many,kumar2020explicitly,
PhysRevX.10.041043,yuwono2024relativistic,windom2024new,gururangan2025extension,madriaga2025pair}, and Green's function theory \cite{van2006quasiparticle,deslippe2012berkeleygw,reeves2025performance}. A significant effort has also been expended to adapt these methodologies as solvers for embedding methods driven by various design principles
\cite{lan2015communication,claudino2019automatic,he2022second,chang2024downfolding,peng2024integrating,shee2024static,park2024tensor,weisburn2025multiscale}. 
%Moreover, as illustrated in several  examples from the last decade, some of these many-body methodologies can be integrated to amplify the accuracy of their predictions (see Refs.\cite{}).   
 
The development of computer implementations for these methods, and their widespread application to realistic problems, requires addressing several challenges arising from the algebraic complexity of the underlying formulations and their polynomially scaling numerical overhead. To manage the algebraic complexity - particularly in coupled-cluster (CC) approaches involving hundreds or even thousands of tensor contractions - the use of symbolic tools for deriving and optimizing tensor expressions is essential for any sustainable development effort.
The symbolic algebra systems that implement Wick's theorem speed up the implementation process and eliminate error-prone and time-consuming hand code effort. In the last decades, the so-called Domain-Specific Languages have become an inextricable element in developing electronic structure methods. 
For example, Tensor Contraction Engine (TCE) \cite{hirata2003tensor}, WICK\&D \cite{evangelista2022automatic}, Symbolic Generator (SymGen) \cite{bylaska2024electronic}, $p^{\dagger}q$ Package \cite{rubin2021p}, or  Sparse Tensor Networks (CoNST) \cite{raje2024const}
supported by various parallel tensor contraction libraries such as Cyclop Tensor Framework (CTF) \cite{solomonik2014massively}, TiledArrays \cite{calvin2019tiledarray,calvin2015scalable}, Tensor Algebra for Many Body Methods (TAMM) \cite{mutlu2023tamm}, 
%or Code Generator for Sparse Tensor Networks (CoNST) \cite{raje2024const} 
provide a complete conduit  from generating second quantized expression, their optimization, and parallel code generation  to the execution on the exascale architectures. 

In recent years, we witnessed activities aimed at the development of scalable implementations of the DMRG methodology \cite{molmps,zhai2021low,zhai2023block2,ganahl2023density,menczer2024parallel}, which resulted in a significant change in the system size limit tractable in DMGR simulations. 
%This numerical effectiveness has recently been illustrated in the  DMRG simulations of strongly correlated systems utilizing compute platforms based on the NVIDIA DGX-H100 GPU. 
While DMRG can yield meaningful results regarding the ordering of electronic states or energy differences between states of various multiplicities in many applications, its focus solely on recovering static correlation effects may adversely impact prediction quality, particularly in situations where the structure of the states of interest strongly depends on the geometry of the chemical system. For this reason, several formulations have been developed to integrate DMRG with perturbation theory and coupled cluster formalisms to address these issues \cite{neuscamman_2010_irpc,Roemelt2016, Freitag2017,sharma_2014c,Veis2016,Beran2021}. 
In Refs. \onlinecite{bauman2019downfolding,downfolding2020t,bauman2022coupled,bauman2025density}, we reported a new algorithm for bringing the missing dynamical correlation effects by employing the effective Hamiltonians corresponding to the Hermitian variant of the CC downfolding \cite{bauman2019downfolding,downfolding2020t}. 
%In this hybrid approach (or hybrid workflow), one first performs CC  calculations based on the double unitary CC Ansatz (DUCC) \cite{bauman2019downfolding,downfolding2020t} to produce a compact many-body representation of the effective (downfolded) Hamiltonian in the relevant active space and then uses effective Hamiltonians, which encodes in its many-body form dynamical correlation effects,  in the context of the DMRG simulations. 
In this hybrid approach (or hybrid workflow), one first performs CC  calculations based on the double unitary CC Ansatz (DUCC) \cite{bauman2019downfolding,downfolding2020t} to produce a compact many-body representation of the effective (downfolded) Hamiltonian in the relevant active space and then use DMRG method to optimize the wave function in the MPS form. 
%effective Hamiltonians, which encodes in its many-body form dynamical correlation effects,  
%in the context of the DMRG simulations. 
We demonstrated the effectiveness of this combined approach (henceforth referred to as the DMRG-DUCC workflow) in reproducing missing dynamical correlation effects \cite{bauman2025density}. Since the DMRG-DUCC workflow contains computational components characterized by significantly different numerical overheads, the utilization of hardware that provides optimal execution environments may be beneficial both from the point of view of time-to-solution and closely related parallel performance as well as the energy/cost footprint of scientific simulations. In this paper, we address this critical issue by combining Compute Express Link (CXL) memory technology developed by Micron for executing memory capacity intensive simulations for effective Hamiltonians, and Microsoft's Azure Quantum Elements (AQE) cloud computing environment for executing more compute intensive DMRG simulations.  

The paper is organized as follows: In Section II, we provide a brief discussion of CC downfolding and DMRG methodologies. Sections III and IV focus on the software components, numerical scaling, and integrated hardware specifications. We also introduce new features of the CC downfolding module that enable users to work with any active space (in the previous application, the design of the CC downfolding module was limited to cases where all occupied orbitals are active). Section V discusses the performance of the TAMM CCSD and CC downfolding implementations available in the ExaChem suite of parallel codes \cite{panyala2023exachem}. Finally, in Section VI, we present the results of the DMRG-DUCC formalism in applications to several strongly correlated molecular systems.
%review OK

\section{Theory}

In this section, we briefly review the theoretical components of the DMRG-DUCC workflow. We also discuss the computational infrastructure for their parallel execution. More details can be found in Refs.~\onlinecite{molmps,mutlu2023tamm,panyala2023exachem,bylaska2024electronic,bauman2025density}.

\subsection{Hermitian CC Downfolding}
The compression of the dimensionality of the quantum problem to the relevant active space has been and continues to be an active area of development in quantum chemistry.
\iffalse
Although effective Hamiltonians have been considered in physics and chemistry for decades in the context of multireference formulations, 
\fi
Recent developments (generally referred to as the  CC downfolding approaches) indicate that the effective Hamiltonian theory is also an inherent feature of standard single reference CC (SR-CC ) formulations \cite{cizek66_4256,paldus07,bartlett_rmp}, allowing for an alternative way of calculating SR-CC ground-state energies (see Refs. \onlinecite{kowalski2018properties,kowalski2021dimensionality,kowalski2023sub}). 
The Hermitian extension of the CC downfolding,\cite{bauman2019downfolding,bauman2022coupled}  which utilizes the so-called double unitary coupled cluster (DUCC) Ansatz, has been introduced in the context of the quantum computing applications to adjust the dimensionality of quantum problem to the available quantum resources.
(for the review of these formulations, see Refs.\cite{metcalf2020resource,doublec2022,kowalski2025resource}). 
In analogy to the non-Hermitian SR-CC case \cite{kowalski2018properties}, the DUCC Ansatz leads to the many-body form of the Hermitian effective Hamiltonian, $H^{\rm eff}$, defined in the active space 
\begin{equation}
    H^{\rm eff} = (P+Q_{\rm int})
    e^{-\sigma_{\rm ext}} H
    e^{\sigma_{\rm ext}}
    (P+Q_{\rm int})
    \label{ducc1}
\end{equation}
%GOK
where $\sigma_{\rm ext}$ is the so-called external anti-Hermitian cluster operator, $P$ is the projection operator onto the reference function $|\Phi\rangle$ (usually chosen as a Hartree-Fock (HF) Slater determinant), and $Q_{\rm int}$ is the projection operator onto excited configurations (with respect to the $|\Phi\rangle$ determinant) belonging to complete active space (CAS) of interest.
%review OK
\iffalse
Once the approximate form of the amplitudes defining $\sigma_{\rm ext}$ operator can be determined then the lowest eigenvalue $E$ of the Hermitian problem 
\begin{equation}
     H^{\rm eff}|\Psi_{\rm act} \rangle = E |\Psi_{\rm act} \rangle\;,
     \label{ducc2}
\end{equation}
approximates the exact ground-state energy of the system obtained in the entire Hilbert  space for a given basis set.  In particular, the above form of the effective Hamiltonian encapsulates the out-of-active-space (or dynamical) correlation effects. 
The eigenvector $|\Psi_{\rm act} \rangle$ can be used to determine the so-called internal cluster operator $\sigma_{\rm int}$ defined within the active space. The above partitioning of the wave function parameters into internal and external cluster operators is inspired by the partitioning of the cluster operators introduced in single reference active-space CC methods \cite{pnl93} to select higher-rank excitations.
\fi
%GOK

In practical applications, the construction of a second quantized representation of $H^{\rm eff}$ is associated with several approximations. First, the $\sigma_{\rm ext}$ operator is approximated in the unitary CC (UCC) form
\begin{equation}
    \sigma_{\rm ext} = T_{\rm ext} - T_{\rm ext}^{\dagger} \;,
    \label{ducc3}
\end{equation}
where $T_{\rm ext}$ is the external part of the  standard CC cluster operator defined by cluster amplitudes that carry at least one  inactive spin-orbital index. 
(the partitioning of the SR-CC cluster operator into the internal and external parts originates in the active-space SR-CC theory, see Ref. \onlinecite{pnl93}). 
For practical reasons, in our approach, the  $T_{\rm ext}$ is approximated by the external part of the CCSD cluster operator. Second, the expansion (\ref{ducc1}) is non-terminating; therefore, we use the finite-rank commutator expansion stemming from the Baker–Campbell–Hausdorff formula. 
\iffalse
, i.e., 
% GOK
%\begin{widetext}
\begin{equation}
   H^{\rm eff} \simeq 
     (P+Q_{\rm int})
    (H +  \sum_{i=1}^{l}  \frac{1}{i!}[
 \ldots [H,\sigma_{\rm ext}],\ldots ],\sigma_{\rm ext}]_i)
    (P+Q_{\rm int})
     \label{ducc4}
\end{equation}
%\end{widetext}
\fi
Due to the high cost of the numerical evaluation of higher-rank commutator contributions, in this paper, we include all double commutators with triple-commutator terms stemming from the Fock operator for the perturbative consistency (see Ref. \onlinecite{doublec2022} for the detailed discussion).
Additionally, the rank of many-body effects included in $H^{\rm eff}$ are limited to the one- and two-body interactions:
\begin{equation}
H^{\rm eff} \simeq
\Gamma_0 +
\sum_{pq} g^p_q a_p^{\dagger} a_q + \frac{1}{4} \sum_{p,q,r,s} k^{pq}_{rs} a_p^{\dagger} a_q^{\dagger} a_s a_r \;,
\label{ducc5}
\end{equation}
where $\Gamma_0$ is a scalar, $g^p_q$ and 
$k^{pq}_{rs}$ tensors defining one- and two-body effective interactions, and indices $p$, $q$, $r$, $s$ designate active spin-orbitals.

The development of coupled-cluster downfolding infrastructure has focused on creating an HPC-capable framework optimized for massively parallel, GPU-based architectures to address realistic chemical problems. Early serial, hand-coded prototypes were essential in establishing the hierarchical structure of downfolded Hamiltonians within the double unitary coupled-cluster (DUCC) framework, particularly DUCC(2) and DUCC(3) approximations \cite{doublec2022}.
%(see Appendix~\ref{appendix:a7}). 
These initial implementations assumed full orbital occupation in the active space \cite{doublec2022}.
Scaling these methods for larger systems required addressing two key challenges:
(1) enabling efficient, GPU-accelerated parallel implementations of CC downfolding, and
(2) supporting arbitrary active space selections to accommodate current DMRG implementations. In all calculations we used frozen core approximation, in which we included the core orbitals. In the downfolding algorithm, the occupied orbitals not included in the active space were considered frozen.

This work employs the DUCC(3) approximation, where effective Hamiltonians are defined by single, double, and partial triple commutators (see Ref. \onlinecite{doublec2022}), with external cluster amplitudes from CCSD calculations. Scalar, one-, and two-body components of the effective Hamiltonian were generated using the parallel DUCC implementation in the ExaChem code (\url{https://github.com/ExaChem/exachem}). The construction of DUCC(3) Hamiltonians involves over thousand Hugenholtz-type diagrams, whose derivation was automated using SymGen (\url{https://github.com/npbauman/SymGen}), a symbolic algebra system that translates second-quantized operators into tensor expressions \cite{bylaska2024electronic}.
These expressions are converted into TAMM format \cite{mutlu2023tamm}, allowing seamless integration with ExaChem and efficient parallel execution.

\subsection{DMRG method}

The {Density Matrix Renormalization Group} (DMRG)\cite{schollwock_2005} employs a tensor network representation of the wavefunction — the Matrix Product State (MPS)\cite{Schollwock2011} — which enables efficient and accurate approximation of low-lying eigenstates of a given electronic Hamiltonian, particularly in systems with strong electron correlation.
Originally formulated for low-dimensional quantum lattice models, DMRG has been successfully adapted to quantum chemistry, where it offers a systematically improvable alternative to traditional methods such as {Full Configuration Interaction} (FCI) and {Complete Active Space Self-Consistent Field} (CASSCF).

While FCI provides exact solutions within a given orbital basis, its factorial scaling severely limits its applicability. CASSCF reduces computational demand by partitioning orbitals into inactive, active, and virtual subspaces, yet remains impractical for active spaces beyond $\sim$20 orbitals due to exponential growth of the configuration space. In contrast, DMRG addresses this bottleneck by expressing the many-body wavefunction as a \textit{Matrix Product State} (MPS), enabling efficient exploration of large active spaces:
\begin{eqnarray}
|\Psi\rangle &=&  \sum_{\{s_i\}} c^{s_1 s_2 
\cdots s_N} |s_1 s_2 \cdots s_N\rangle \nonumber \\
&=&
\sum_{\{s_i\}} \sum_{i_1\ldots i_{N-1}} 
A[1]^{s_1}_{i_1} A[2]^{s_2}_{i_1 i_2} 
A[3]^{s_3}_{i_2 i_3}\cdots A[N]^{s_N}_{i_{N-1}} |s_1 s_2 \cdots s_N\rangle,
\label{dmrg1}
\end{eqnarray}
where tensor $A[n]^{s_n}_{i_{n-1}i_n}$ carries three indices: one corresponding to the physical basis ($s_i$), and two auxiliary indices connecting to adjacent tensors ($i_{n-1}$ and $i_n$), often referred to as {virtual bonds} (M).

%The {bond dimension} $M$ is defined as the maximum dimension of these virtual bond indices:
%\begin{equation}
%A_i^{s_i} \in \mathbb{C}^{M_{i-1} \times d \times M_i},
%\label{dmrg2}
%\end{equation}
%where $d$ is the dimension of the local Hilbert space (e.g., $d = 4$ for a spatial orbital), and $M_{i}$ controls the amount of entanglement allowed between site $i$ and the rest of the system.

A larger bond dimension $M$ allows the MPS to capture more entanglement and hence more accurately approximate the true ground-state wavefunction. In the limit $M \to \infty$, the MPS becomes an exact representation of the wavefunction (within the given orbital basis). During DMRG optimization, singular value decomposition (SVD) is employed to truncate the virtual bond dimensions, retaining only the $M$ largest Schmidt coefficients and discarding less significant entanglement contributions. This truncation controls both the computational cost and the accuracy of the method.

DMRG is particularly effective at capturing {static correlation}, which arises in near-degenerate and multireference electronic configurations. However, as with traditional CASSCF-based methods, it does not inherently recover {dynamic correlation} from external space, which is essential for quantitatively accurate results and stems from short-range, high-energy electron interactions.

To address this limitation, we employ here a hybrid DMRG-DUCC  framework. In this approach, DMRG is used to accurately describe the static correlation within a strongly correlated active space, while the DUCC formalism systematically integrates out the external (dynamically correlated) degrees of freedom. Specifically, DUCC constructs an effective Hamiltonian for the active space by applying a unitary transformation that decouples it from the high-energy virtual subspace, leading to a renormalized, downfolded Hamiltonian that captures the essential dynamical contributions. This formulation avoids perturbative approximations and maintains a rigorous wavefunction-based foundation, making it especially well-suited for strongly correlated systems where standard perturbation theories (e.g., CASPT2 \cite{andersson1992second}, NEVPT2 \cite{angeli2001introduction}) may fail or break down.

The DMRG-DUCC combination should offer a multiscale correlation treatment: DMRG handles the low-energy, entangled sector with controlled precision, while DUCC accounts for dynamic correlation effects through a systematic, non-perturbative renormalization of the Hamiltonian. This strategy enables obtaining highly accurate properties in large molecular systems with extended active spaces that are intractable using conventional methods.

All DMRG calculations in this work were carried out using our in-house software package, \textsc{MOLMPS}\cite{molmps}, which implements a robust massively parallel DMRG solver with orbital ordering based on entanglement measures, and the interface with external downfolding framework.

\section{Software components and numerical scaling}
In our workflow, the DUCC part requires two essential steps:
\begin{itemize}
\item Providing an approximate form of the $T_{\rm ext}$ operator (Eq.~(\ref{ducc3})). In the present study, this operator is approximated using the external components of single and double cluster operator obtained from CCSD calculations.

\item Constructing an approximate many-body form of the effective Hamiltonian within the relevant active space. In our simulations of downfolded Hamiltonians, we focus on the scalar term as well as the one- and two-body interactions.
\end{itemize}
Both computational components are implemented in the ExaChem code ((\url{https://github.com/ExaChem/exachem}). While these methodologies scale as ${\cal N}^6$  for CCSD and $ N_{\rm act}^3\times{\cal N}^3$  for CC downfolding ($N_{\rm act}$ and ${\cal N}$ stand for the number of active orbitals and system-size, respectively), the DUCC part requires a significantly larger number of diagrams compared to the CCSD formalism. Moreover, DUCC-based downfolding involves a more complex structure of recursive intermediates, which are essential for reducing the computational overhead of the CC downfolding calculations.

In the specific calculations we performed the DMRG calculations have been less expensive compared to the composite CCSD and DUCC simulations, but in general this strongly depends on the complexity of the correlation structure inside the computed active space. Typically, the associated overhead (for a single sweep) is proportional to 
$M^3\times N_{\rm act}^3 + M^2 \times N_{\rm act}^4$ where $M$ stands for the bond dimension. 
%{\color{red} Jirka: we need more detailed discussion here}

%revision OK

\section{Hardware specifications }
%The goal of is provide hardware which is adequate to the numerical task and provides 
The goal of this section is to describe the hardware used in the DMRG-DUCC workflow. The DUCC simulations were performed on the CXL system, while the DMRG calculations utilized the AQE cloud computing infrastructure.

\subsection{Micron CXL 2.0 Memory Lake System}
With the arrival of CXL hosts, CXL 2.0 switches, and CXL 2.0 memory modules, it will be possible to build servers with memory capacities in the 10TB to 100TB range.
Historically, scale-up servers have been the only way to achieve these levels of memory capacity.
For example, HPE’s Superdome-Flex 280 server\cite{HPESuperDome} scales up to 28TB. These scale up servers are expensive, have limited GPU support, and have never been able to scale to 100TB.
Micron is researching and prototyping CXL clusters built using the XC50256 CXL 2.0 switch manufactured by XCONN Technologies\cite{xconntechProductXconn}. The XCONN switch is a 32 ported switch capable of supporting two levels of CXL switching.
This switch enables the industry to stand up servers capable of addressing 33TB with just one level of CXL switching and 132TB with two levels of CXL switching. Larger capacities would be possible with higher capacity CXL memory modules.

\begin{table}[]
\begin{tabular}{|l|l|l|l|l|}
\hline
Levels of CXL Switching                                                           & One                                                                            & Two                                       \\ \hline
\begin{tabular}[c]{@{}l@{}}Max CXL Capacity\\ (256 GB Modules)\end{tabular} & 33TB                                                                           & 132TB                                     \\ \hline
Hardware Technology Need                                                          & \begin{tabular}[c]{@{}l@{}}CXL Switch,\\CXL repeater cards\end{tabular} & \begin{tabular}[c]{@{}l@{}}CXL Fiber Optic\\ Active Optical Cables\end{tabular} \\ \hline
CXL Idle Latency                                                                  & 520ns                                                                          & 810ns                                     \\ \hline
Max Usable CXL Bandwidth                                                          & 132 + 132 GB/s/Socket                                                          & 88 + 88 GB/s/Socket                       \\ \hline
Number of Servers (Hosts)                                                         & One to Four                                                                    & One to Four                               \\ \hline
\end{tabular}
\end{table}

\FloatBarrier

\subsubsection{Prototype Cluster}
The experiments in this paper were run on a 7.5TB shared memory prototype cluster with the following specs:
\begin{center}
\begin{itemize}
  \item One XCONN XC50256 switch
  \item Four Intel Xeon Emerald Rapids CXL hosts, with 1TB DRAM
  \item Fifteen FPGA based 512 GB CXL modules.  Each module is a PCIe card with four 128GB DDR4 DIMMs.
\end{itemize}
\end{center}

\begin{figure}
    \includegraphics[width=0.95\textwidth]{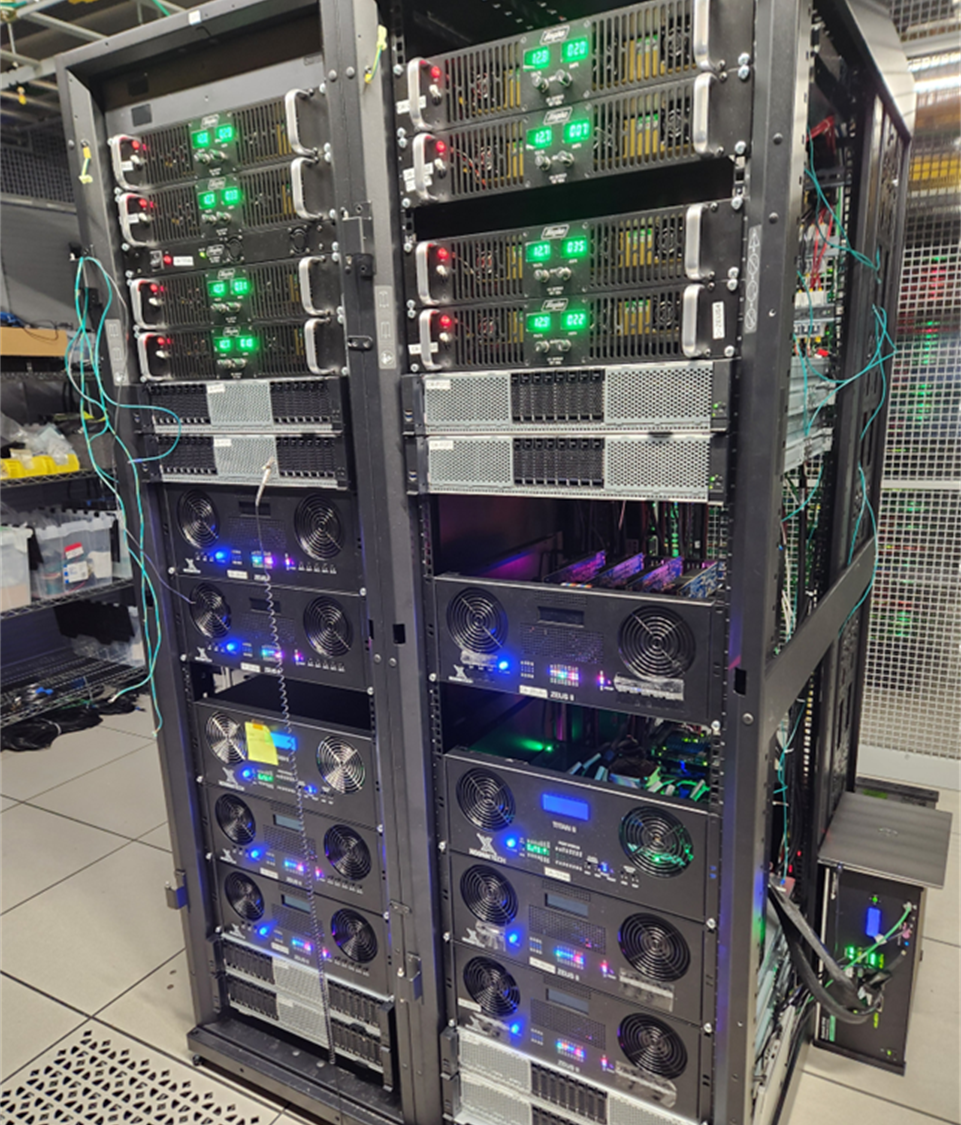}
    \caption{CreteA and CreteB clusters (each using 1 rack of equipment) built for PNNL under the AMAIS project. There are four power suplies at the top, two hosts in the middle and the bottom, 1 large chassis for the CXL switch, and 4 large chassis with multiple FPGA boards. }
    \label{fig:prototype-cluster}
\end{figure}

\subsubsection{Multi Host Support}
Multi-host hardware cache coherency is not supported by any CXL 2.0 switch or processor.
Multi-host sharing of CXL 2.0 devices using software cache coherency is NOT in the CXL 2.0 spec, but IS supported by the XCONN switch and Intel Xeon Emerald Rapids CXL hosts.
Software cache coherency will be required for multi-host sharing until hardware cache coherency is enabled by future CXL 3.1 processors.

To use shared memory across hosts, Micron has developed a proof-of-concept shared memory allocator called Rapid.
When using Rapid, the OS must configure CXL memory modules in Direct Access (DAX) mode.
Rapid maps the CXL memory contiguously in the address space of applications at the same offset.
Rapid also provides a system service called the memserver that handles all allocation requests for the cluster. This manages the global state of which regions of memory have been allocated to which hosts.
Using software coherency semantics, store operations performed on one host will be visible from the other hosts.

This cluster has enabled Micron to prototype multi-host CXL memory sharing and experiment with software cache coherency schemes
(a prototype cluster hosted at PNNL is shown in Fig. \ref{fig:prototype-cluster}).
The experiments pointed to a specific sequence of flush and fence calls needed for both producer and consumer that is needed to ensure data is entirely flushed to CXL media and readable by other hosts.
Those intrinsics have been packaged into a small software coherency library that is used by Rapid and applications we have ported to use fabric attached memory.
Rapid re-exports these software coherency functions through the library API so it is simpler for users to insert the correct memory operations.

%%\subsection{AQE cloud computing environment}
%\include{aqe}
%%%\input{aqe.tex}
\subsection{AQE cloud computing environment}
%% TODO: AQE Hardware specification

Azure Quantum Elements (AQE or Elements) was created to accelerate scientific discoveries by combining the power of cloud high-performance computing, artificial intelligence (AI), and emerging quantum capabilities. The AQE platform offers a hybrid computing environment that saves valuable time and resources in the scientific process.  The purpose of AQE is not to replace traditional laboratory experimentation, but to supplement it by offering a powerful computing platform. The AQE platform helps researchers in chemistry and materials science narrow the vast space of molecules and materials by using AI screening, HPC simulations including density functional theory and molecular dynamics so that limited laboratory resources can be spent exploring only the most promising candidates. By bringing computing to the forefront of scientific research, the AQE platform has the potential to accelerate discoveries in multiple fields and industries including transportation, sustainability, and pharmaceuticals.
Elements offer a wide variety of hardware for HPC , see Table ~\ref{tab:AQE_slurm},  including utility nodes for scripting, analysis of results, rendering, and visualization. The CPU and GPU nodes are interconnected by InfiniBand making LCF workloads easy to migrate to the Azure cloud, and the heterogeneous nature of the hardware is suitable for diverse workloads and workflows.
%\textcolor{blue}{
Azure Quantum Elements has been a private preview product and its pricing structure varies based on customer workloads and customer requirements. The upcoming Microsoft Discovery platform will subsume and is the next iteration of AQE.
%}

\begin{table}[h]
\centering

\begin{tabular}{|>{\raggedright\arraybackslash}p{0.15\linewidth}|>{\raggedright\arraybackslash}p{0.15\linewidth}|>{\raggedright\arraybackslash}p{0.15\linewidth}|>{\raggedright\arraybackslash}p{0.15\linewidth}|>{\raggedright\arraybackslash}p{0.15\linewidth}|>{\raggedright\arraybackslash}p{0.15\linewidth}|>{\raggedright\arraybackslash}p{0.15\linewidth}|>{\raggedright\arraybackslash}p{0.15\linewidth}|>{\raggedright\arraybackslash}p{0.15\linewidth}|>{\raggedright\arraybackslash}p{0.15\linewidth}|} \hline
VM & vCPU & Memory (GiB) & GPU & InfiniBand (Gb$/$s)  & Max network bandwidth (Mbps) \\ \hline
D8 v5 & 8 & 32 & NA & NA & 12,500 \\ \hline
NV36 v5 & 36 & 880 & 1 A10 24 GB GPU & NA & 80,000  \\ \hline
HB120 v2& 120 & 456 & NA & 200  & 50,000  \\ \hline
ND40 v2& 40 & 672 & 8 V100 32 GB GPUs (NVLink)& 100 & 24,000 \\ \hline
NC96 v4& 96 & 880 & 4 A100 80 GB GPUs (NVLink)& NA & 80,000 \\ \hline
ND96 v5& 96 & 1900 & 8 H100 80 GB GPUs (NVLink)& 400 & 80,000 \\ \hline
NCads v5& 80 & 640 & 2 H100 80 GB GPUs (NVLink)& NA & 80,000 \\ \hline

\end{tabular}
\caption{AQE Slurm Partition Hardware Specifications as of July 2025.}
\label{tab:AQE_slurm}
\end{table}

%% TODO: CloudOps and corresponding architecture for integration between PNNL and AQE Azure tenant

%% TODO: Evolution of the AQE platform into the SuperComputing sub-system of Azure Discovery.

% revision OK

\section{Performance Analysis}

\subsection{Porting TAMM to Fabric Attached Memory (FAM)}
A CXL Fabric Attached Memory Pool enables allocating one large tensor across the fabric that all compute nodes can share and directly reference for their portion of the calculation.
TAMM allocates tensors using GlobalArrays (GA), a library written by PNNL that enables aggregating the RAM of hosts in a network based distributed system to create “shared memory” arrays that span the system.
The default GA backend is an MPI based implementation that handles the network copies as compute nodes reference remote memory.

\begin{figure}
    \includegraphics[width=0.8\textwidth]{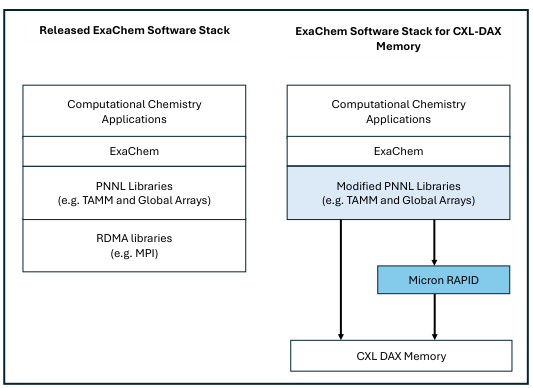}
    \caption{Updating the ExaChem software stack to support CXL shared memory}
    \label{fig:software-stack}
\end{figure}

To port TAMM to CXL shared memory(Figure~\ref{fig:software-stack}), Micron used its Rapid library that presents CXL fabric attached memory as a shared virtual address space to applications.
Micron modified TAMM's tensor memory allocation function to call the Rapid memory allocator instead of the GA allocator.   Allocation is performed only by rank 0 and the pointer to the allocated memory is broadcast to all other ranks.   Each rank maintains a copy of the pointer associated with each tensor.
Tensor memory can then be accessed directly by any rank on any host.   No temporary buffers or copying are required.

\begin{figure}
    \includegraphics[width=0.8\textwidth]{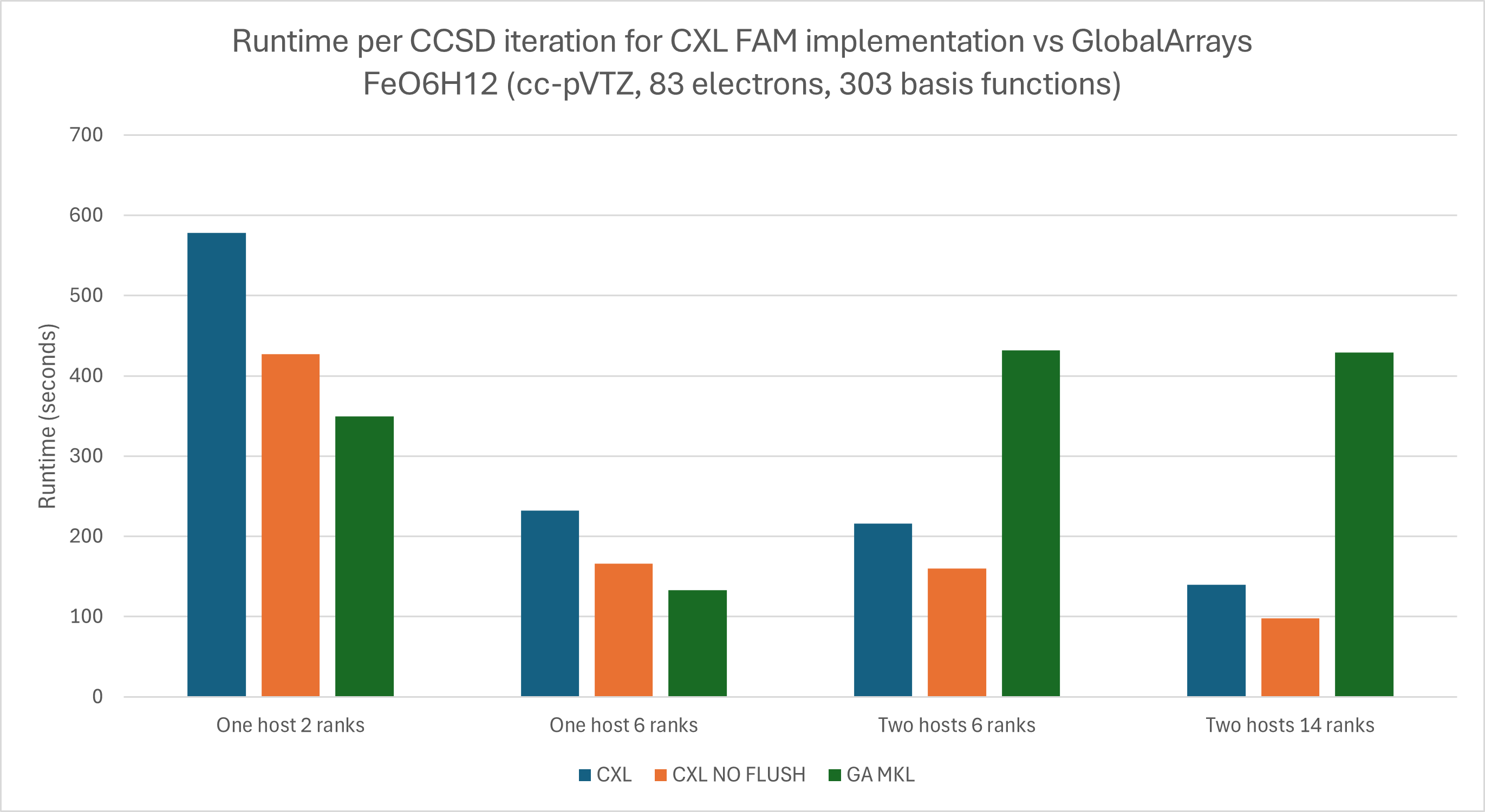}
    \caption{Comparison of CCSD per-iteration runtimes using [Fe$(\rm H_2O)_6$](3+) with the cc-pVTZ basis set (no frozen core): CXL FAM based implementation vs Global Arrays (GA) distributed memory implementation}
    \label{fig:testccsd-perf}
\end{figure}

Figure~\ref{fig:testccsd-perf} shows a comparison of the runtime performance running a TAMM CCSD test application using CXL (with and without coherency flushing) versus the original GA version. For the benchmark we have chosen [Fe$(\rm H_2O)_6$](3+) system \cite{Rado2019} in the cc-pVTZ basis set, yielding a total of 303 basis functions. In this case, the frozen-core approximation was not employed and all electrons were explicitly correlated. When running on a single host, performance of the GA version is better than that of the CXL versions.  GA is using local host memory which has higher performance than the CXL memory prototype.   When running on two hosts, the CXL versions are faster because they do not incur the overhead of copying data between hosts.    The CXL versions are also able to benefit from an increased number of ranks when running on two hosts.

% revision OK

\section{Results}

We have selected two test cases to demonstrate the role of the downfolding technique in describing electronic systems. The first is Fe–nitrosyl isomerization, which allows us to compare ground-state energies with and without the inclusion of DUCC(3). This system has been studied recently,\cite{Beran2023} and it has been shown that incorporating dynamical correlation is essential to correctly predict the energy ordering of the isomers. 

\begin{figure}[h]
\centering
\includegraphics[width=0.75\textwidth]{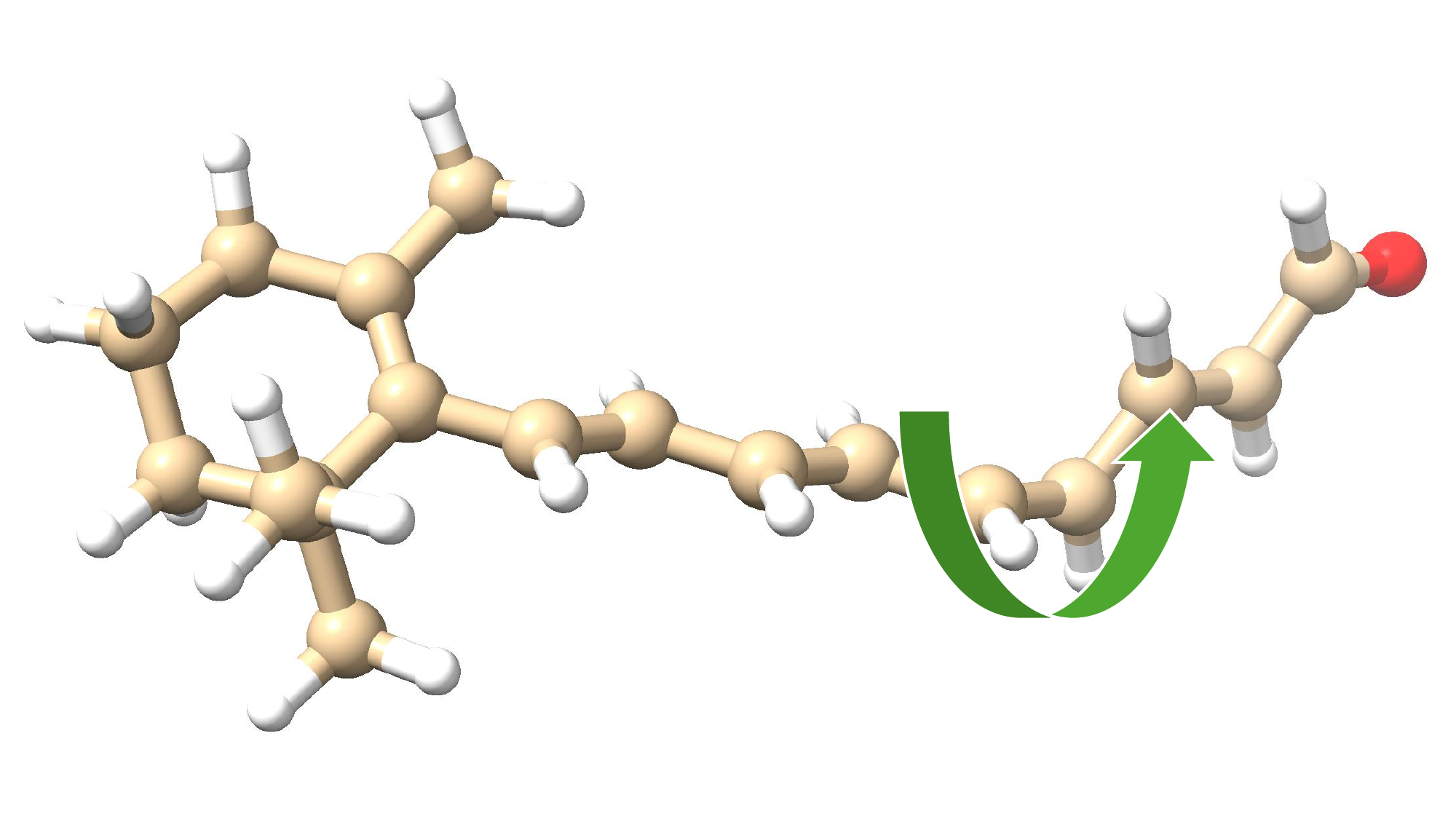}
\caption{The retinal cis-trans isomerization.}
\label{fig:retinal_iso}
\end{figure}

The second system is retinal, which is a co-factor in rhodopsin. The absorption of a photon is connected with cis-trans isomerization of retinal (Fig.\ref{fig:retinal_iso}). It represents a challenging system, which involves changing strength of the static correlation effects.\cite{Claudino2019,Migani2004} While for the rotation angle $\alpha=0$ or $\alpha=180$ the molecule has a closed-shell single-reference character, the system with the $\alpha=90$ twisted double bond is in the intersection point, where the state exhibits multireference open-shell character. It allows us to test limits of the DUCC-based approach, which has been recently developed to treat both dynamical and statical correlations.\\

%The second system is the charge-transfer complex formed by tetrathiafulvalene (TTF) and tetracyanoquinodimethane (TCNQ). The charge-transfer (CT) state is stable, and in its crystalline form, the material is known as an organic semiconductor. Individually, TTF and TCNQ are symmetric molecules and insulators; however, their stacked crystal structure exhibits strong charge transfer. Our model consists of a single TTF–TCNQ unit, for which we calculated the electron density of the complex, as well as those of the individual TTF and TCNQ molecules. \\

\begin{figure}[!ht]
  \begin{center}
    \includegraphics[width=450pt]{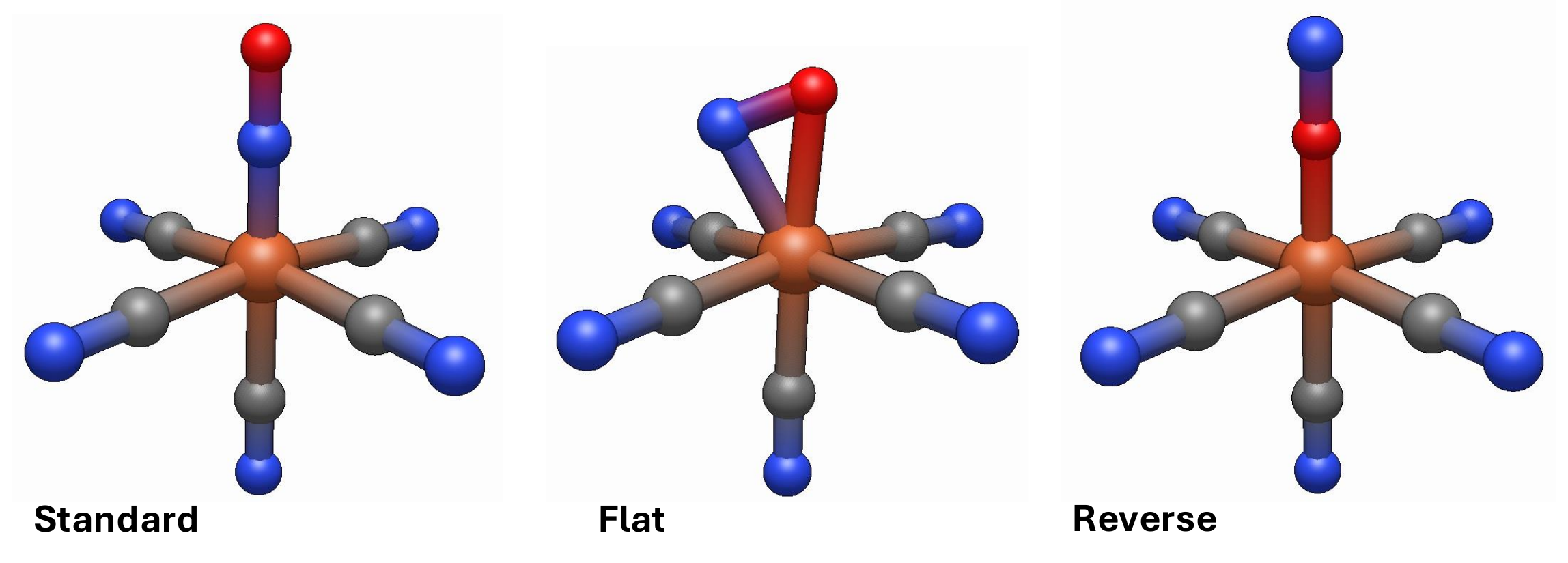}
    \caption{Isomers of the iron-nitrosyl complex. The orange color is for iron, gray for carbon, blue for nitrogen and red for oxygen.}
    \label{fenitro1}
  \end{center}
\end{figure}

To assess the relative stability of the Standard, Flat, and Reverse isomers of the Fe-nitrosyl complex - see Fig.~\ref{fenitro1}, we computed the lowest singlet-state energies using both bare DMRG and DMRG-DUCC methods. 

\begin{figure}[!ht]
  \begin{center}
    \includegraphics[width=450pt]{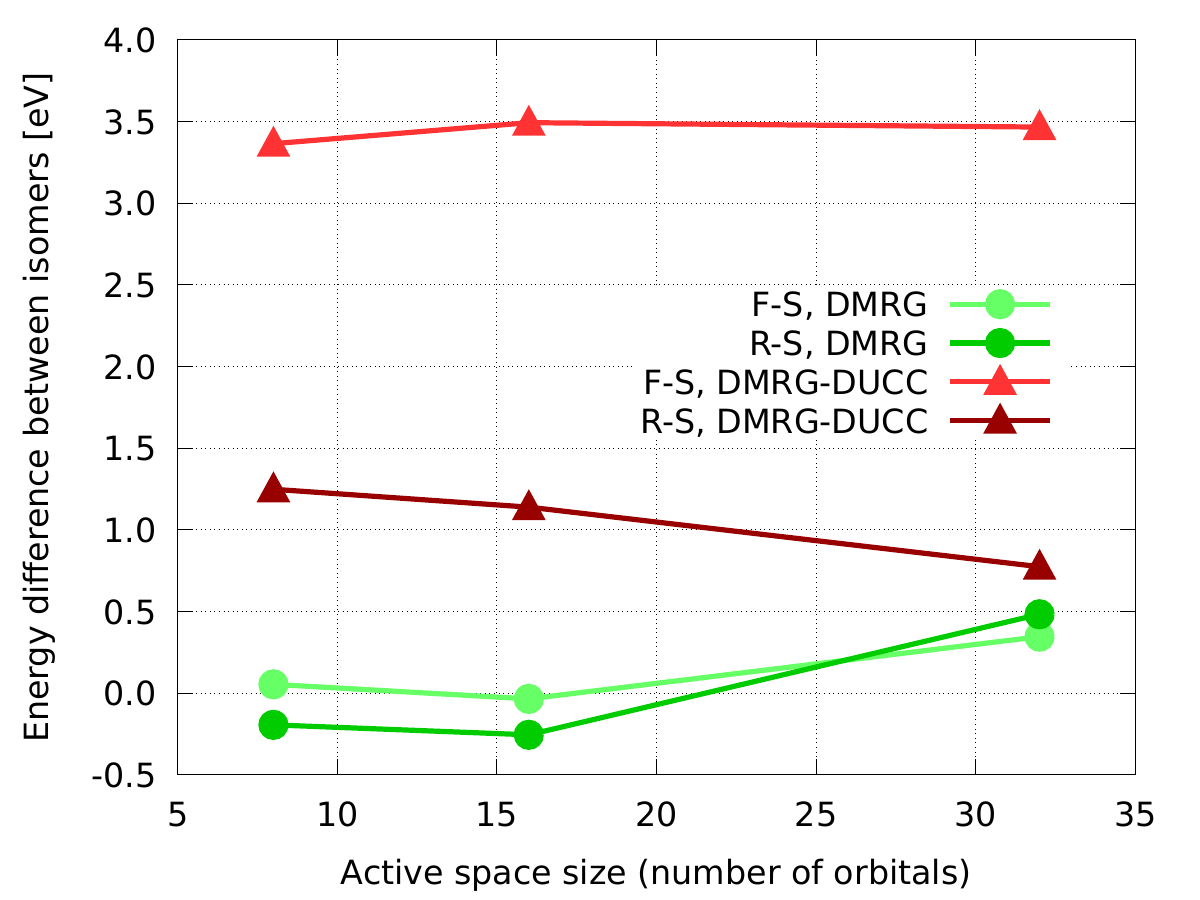}
    \caption{The energy differences (in eV) between Flat - Standard (F-S) and Reverse - Standard (R-S) isomers, computed using DMRG and DMRG-DUCC methods. The used active space is CAS(8,8), CAS(16,16) and CAS(32,32).}
    \label{fenitro2}
  \end{center}
\end{figure}

We utilized the cc-pVTZ basis set\cite{dunning1989gaussian} and considered three active spaces to probe the effect of active space enlargement on the computed energy gaps: (8,8), (16,16), and (32,32). The results are shown in Fig.~\ref{fenitro2}. The bare DMRG results exhibit limited consistency across active spaces: the Flat–Standard energy gap fluctuates in sign and magnitude (0.054 eV, –0.035 eV, and 0.346 eV for (8,8), (16,16), and (32,32), respectively), while the Reverse–Standard gap remains consistently negative or small (–0.194 eV, –0.255 eV, and 0.483 eV). In contrast, the DMRG-DUCC energy differences demonstrate systematic convergence and are substantially larger in magnitude due to the inclusion of dynamic correlation. Specifically, the Flat–Standard gap rises from 3.36 eV to 3.49 eV and stabilizes at 3.47 eV, whereas the Reverse–Standard gap decreases steadily from 1.25 eV to 0.77 eV across the same active spaces. These results highlight the importance of incorporating dynamic correlation via downfolding and demonstrate the stabilizing role of the DUCC treatment on relative energetics among Fe–nitrosyl isomers. \\

%\begin{figure}[!ht]
 % \begin{center}
  %  \includegraphics[width=450pt]{ttftcnq.pdf}
  %  \caption{The structure of the TTF-TCNQ complex, single unit.}
  %  \label{ttftcnq}
  %\end{center}
%\end{figure}

%\section{Retinal isomerization}

%Retinal is a co-factor in rhodopsin. The absorption of a photon is connected with cis-trans isomerization of retinal (Fig.\ref{fig:retinal_iso}). It represents a challenging system, which involves changing strength of the static correlation effects.\cite{} While for the rotation angle $\alpha=0$ or $\alpha=180$ the molecule has closed-shell single-reference character, the system with the $\alpha=90$ twisted double bond is in the intersection point, where the state exhibits multireference open-shell character. It allows us to test limits of the DUCC-based approach, which has been recently developed to treat both dynamical and statical correlations.\\

The geometries of the retinal were taken from the supplementary information of Ref. \onlinecite{Claudino2019}. We employed the CCSD, CASSCF, NEVPT2, DMRG, and DMRG-DUCC methods in the investigation of retinal. The cc-pVDZ basis set\cite{dunning1989gaussian}was used, which consisted of 386 orbitals. For the basic active space, we selected all the p-orbitals responsible for conjugation, further optimized by CASSCF. Those orbitals have been split-localized by the Foster-Boys algorithm.\cite{foster1960canonical} In order to estimate the completeness of the active space,
% AS
we selected an extended active space (26,26), which includes the optimised 12 orbitals, plus 14 orbitals to achieve CAS(26,26). The added orbitals were split-localized separately. \\

\begin{figure}[h]
\centering
\includegraphics[width=0.75\textwidth]{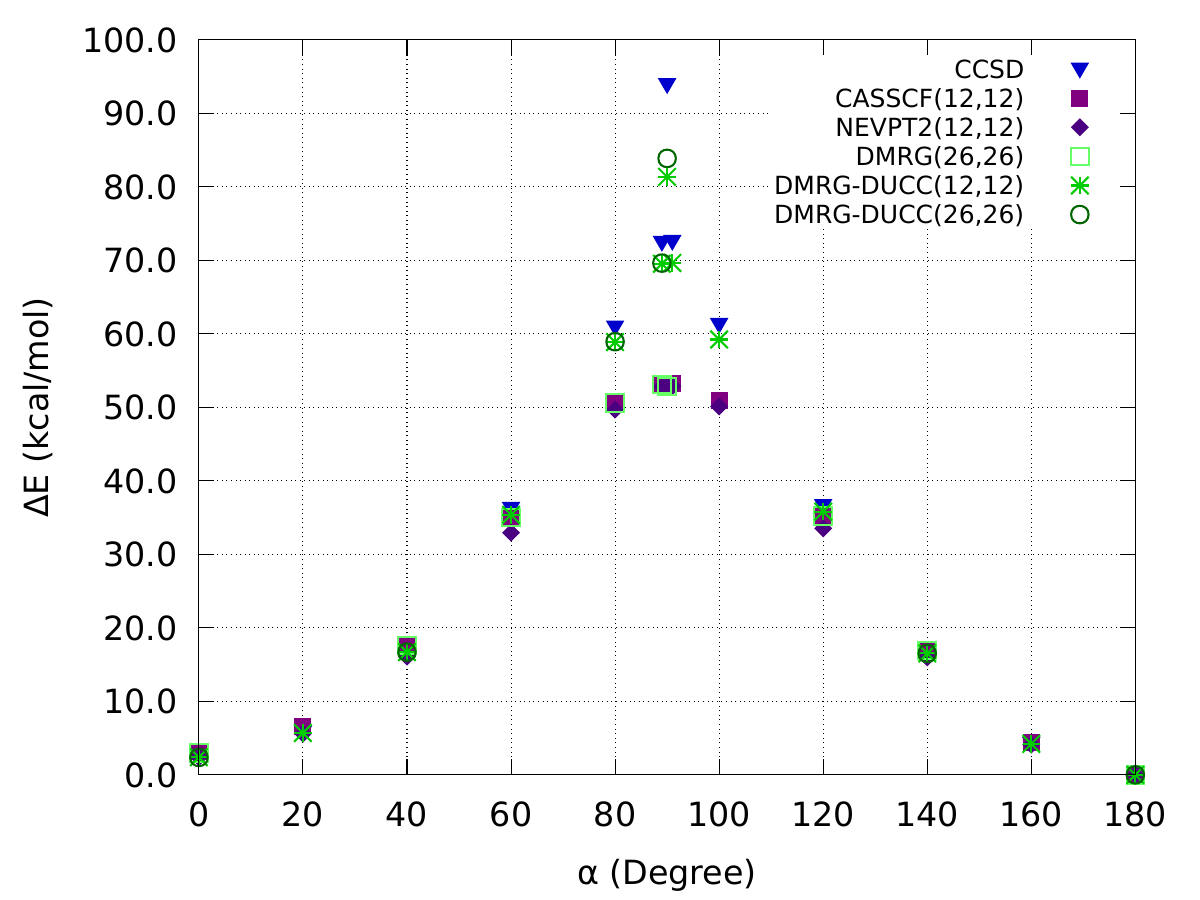}
\caption{The retinal isomerization energy profile in kcal/mol, the rotation angle $\alpha$ is between 0 and 180 degrees. The energy is relative to the $\alpha=180$ case.}
\label{fig:scal}
\end{figure}
The energy profile is in Fig.\ref{fig:scal}, where $\Delta E$ corresponds to the energy difference obtained as $\Delta E=E_{\alpha} - E_{180}$. The lowest energy always corresponds to $\alpha=180$, for $\alpha=0$ the energy is approximately 2.5 - 3 kcal/mol higher. The NEVPT2 and CASSCF(12,12)/DMRG(26,26) methods give a smooth profile, $\Delta E_{\rm NEVPT2}$ and $\Delta E_{\rm CASSCF}$ or $\Delta E_{\rm DMRG}$ are similar (53 kcal/mol), with $\Delta E_{\rm NEVPT2}$ always slightly smaller (within 1 kcal/mol). These results suggest that the energy contribution from dynamical correlation does not change during the rotation, the largest difference between $\Delta E_{\rm CASSCF}$ and $\Delta E_{\rm NEVPT2}$ is 2 kcal/mol. \\      

The CCSD method breaks down for $\alpha=90$ as $\Delta E_{\rm CCSD}=93.9$ kcal/mol (including problematic convergence), $\Delta E$ for $\alpha=89$ or $\alpha=91$ is 72.5 or 72.6 kcal/mol, respectively. This reflects the fact that CCSD does not treat static correlation properly and the correlation energy is underestimated for twisted angles. The DMRG-DUCC methods show similar profile to CCSD, but $\Delta E$ is smaller. When compared to NEVPT2 results, $\Delta E_{\rm CCSD}$ is much larger. When we analyzed the natural occupation numbers $n$ obtained from DMRG calculations, for angles close to 90 the $n_{\rm HONO}$ and $n_{\rm LUNO}$ is close to 1.0. However, when downfolded Hamiltonian is used in DMRG-DUCC approach, $n_{\rm HONO}$ and $n_{\rm LUNO}$ are approximately 1.6 and 0.4. This can be explained by the fact, that the amplitudes used to construct $\sigma_{\rm ext}$ in Eq \ref{ducc3} are from single-reference CCSD method, so the effective Hamiltonian is strongly biased. Therefore, $\Delta E_{\rm DMRG-DUCC}$ is underestimating the static correlation, even if the initial active space is exactly the same as for bare Hamiltonian DMRG method. 

\begin{figure}[h]
\centering
\includegraphics[width=0.75\textwidth]{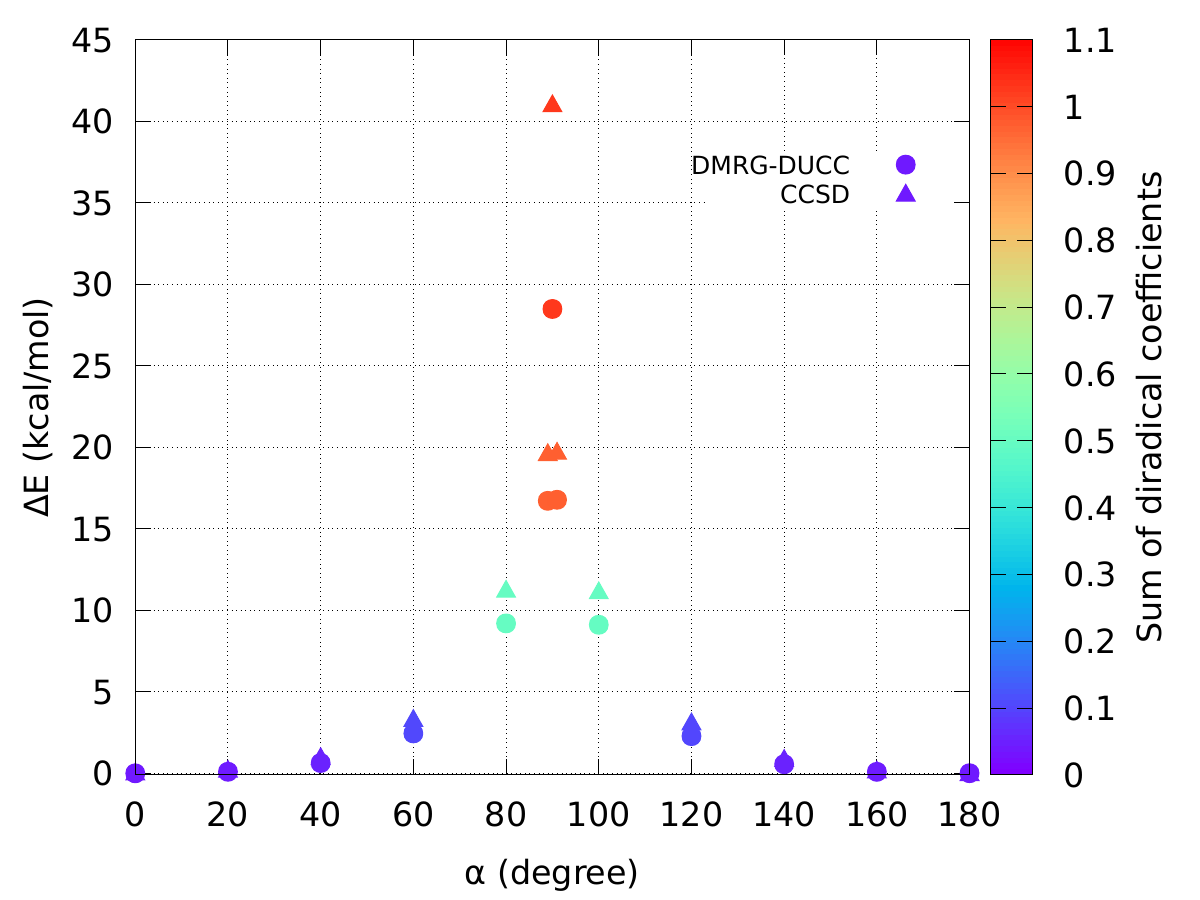}
\caption{The energy difference defined as $\Delta E_{\rm DMRG-DUCC} - \Delta E_{\rm NEVPT2}$, with respect to the sum of diradical coefficients.}
\label{fig:diradical}
\end{figure}
To determine how $\Delta E_{\rm DMRG-DUCC}$ is reliable with respect to the diradical character of the system, we computed the diradical coefficient\cite{Yamaguchi1988}, which define how strong is the open-shell character of the system:
\begin{equation}
Y_{\,i} = 1 - \frac{2 T_i}{1 + T_i^{2}}
\label{eq:diradical_index}
\end{equation}
where $T_i = \tfrac{n_{\text{HONO}-i} - n_{\text{LUNO}+i}}{2}$.
In Fig.\ref{fig:diradical} we plot the difference of $\Delta E$ between DMRG-DUCC and NEVPT2 as $\Delta E_{\rm DMRG-DUCC} - \Delta E_{\rm NEVPT2}$ versus the sum of diradical coefficients $S=\sum_{i\in \rm AS}Y_{\,i}$, where AS is a set of the actual active orbitals. The results show that if $S<0.1$, the difference $\Delta E_{\rm DMRG-DUCC} - \Delta E_{\rm NEVPT2}$ is up to 2.5 kcal/mol, but for larger $S$, the DMRG-DUCC approach underestimates the static correlation effect significantly.

\section{Conclusion}

In this paper, we outlined the DMRG-DUCC workflow that connects two emerging computational technologies: Micron’s CXL technology and the AQE cloud computing platform. While the latter is employed for DMRG applications, the former is better suited for CCSD and DUCC simulations, which are more memory- and computation-intensive. We demonstrated the clear advantages of using the RAPID memory allocator in multi-host executions. Azure Quantum Elements provided fast access to the computational resources required to perform a large number of DMRG calculations.

In our studies of (1) iron–nitrosyl and (2) retinal cis–trans isomerization, we examined the role of dynamical correlation effects. For the first system, the DUCC Hamiltonians provided systematic convergence and quantified the impact of external dynamical correlation effects at the CCSD level. For the second system, the DUCC formalism significantly improved upon the poor performance of the CCSD method compared to the NEVPT2 approach.

We believe the DUCC formalism offers a clear path toward further improving the quality of downfolded Hamiltonians, for example, through the inclusion of higher-rank many-body components (see Ref. \onlinecite{singh2025qubit}) and more accurate forms of the approximate $\sigma_{\rm ext}$ operator. A natural next step toward this goal is the incorporation of three-body effects into $\sigma_{\rm ext}$.

Additionally, CXL architectures and the AQE cloud environment can be leveraged in hybrid classical/quantum workflows, where classical computational components generate downfolded Hamiltonians for quantum simulations of realistic systems \cite{bauman2025coupled}.

\section{Acknowledgement}
This material is based upon work supported by the ``Transferring exascale computational chemistry to cloud computing environment and emerging hardware technologies (TEC$^4$)''  project, which is funded by the U.S. Department of Energy, Office of Science, Office of Basic Energy Sciences, the Division of Chemical Sciences, Geosciences, and Biosciences (under FWP 82037).  
 This work used resources from the Pacific Northwest National Laboratory (PNNL).
PNNL is operated by Battelle for the U.S. Department of Energy under Contract DE-AC05-76RL01830. LV and JB also acknowledge financial support from the Czech Science Foundation (grant no. 23-05486S), the Ministry of
Education, Youth and Sports of the Czech Republic through the e-INFRA CZ (ID:90254), and the Advanced Multiscale Materials for Key Enabling Technologies project, supported by the Ministry of Education, Youth, and Sports of the Czech Republic. Project No. CZ.02.01.01/00/22\_008/0004558, Co-funded by the European Union.

\bibliography{references}

\end{document}